\documentclass[aps,prb,twocolumn,groupedaddress,showpacs]{revtex4-1}
\usepackage{graphics}
\usepackage{graphicx}
\usepackage{amsmath}
\usepackage{amssymb}
\usepackage{amsthm}
\usepackage{dsfont}
\usepackage{xcolor} 
\begin{document}
\title{Study of the superconducting to normal transition}
\author{Jacob Szeftel$^1$}
\email[corresponding author :\quad]{jszeftel@lpqm.ens-cachan.fr}
\author{Nicolas Sandeau$^2$}
\author{Michel Abou Ghantous$^3$}
\affiliation{$^1$Laboratoire de Photonique Quantique et Moléculaire, UMR 8537, Ecole Normale Supérieure de Paris-Saclay, CentraleSupélec, CNRS, Université Paris-Saclay, 94235 Cachan, France}
\affiliation{$^2$Aix Marseille Univ, CNRS, Centrale Marseille, Institut Fresnel, F-13013 Marseille, France}
\affiliation{$^3$American University of Technology, AUT Halat, Highway, Lebanon}
\date{\today}
\begin{abstract}
A model, based on classical mechanics and thermodynamics and valid for \textit{all} superconductors, is devised to investigate the properties of the \textit{current-driven}, superconducting to normal transition. This process is shown to be reversible. An original derivation of the BCS variational procedure is given. Two different critical temperatures are introduced. The temperature dependence of the critical current is worked out and found to agree with observation. The peculiar transport properties of high-$T_c$ compounds in the normal state and old magnetoelastic data are also interpreted within this framework. A novel experiment is proposed to check the relevance of this analysis to high-$T_c$ superconductivity.  
\end{abstract}
\maketitle
		\section{Introduction}
	Shortly after the discovery of superconductivity, it was realized that applying a growing magnetic field\cite{par,tin,sch,gen} $H$ turns superconducting electrons into normal ones at a critical value $H=H_c$. Besides, this process has been characterized as a reversible first order transition, i.e. decreasing $H$ from $H_c$ down to $0$ brings normal electrons back to the superconducting state. However, this experimental procedure suffers from several drawbacks, when scrutinized from a thermodynamical standpoint :
	\begin{itemize}
		\item 
		because all experiments have been made so far\cite{par,tin,sch,gen,and,rul1,laa,sug,rul,gri} at fixed temperature $T$, the heat exchanged during the transition remains unknown. Likewise, since nobody bothered to measure the work performed by $H$,  the binding energy of the superconducting phase with respect to the normal one $E_b$ could not be assessed with help of the first law of thermodynamics. Meanwhile the formula $E_b=\mu_0H_c^2/2$ was assumed\cite{gor}, with $\mu_0$ being the magnetic permeability of vacuum, and became eventually ubiquitous in textbooks\cite{par,tin,sch,gen}. Unfortunately, a numerical application in case of $Al$ turns out to underestimate\cite{sz2} by ten orders of magnitude the $E_b$ value, deduced from the BCS theory\cite{bcs}; 
		\item 
		due to the Meissner effect\cite{sz2} and the finite \textit{ac} conductivity\cite{sch} in the superconducting state, the current density is spatially inhomogeneous\cite{sz2,sz1,sz3} and there is no one-to-one correspondence between the external magnetic field  and the current distribution inside the sample, so that \textit{qualitative} information only can be achieved from $H$ mediated experiments\cite{par,tin,sch,gen,and,rul1,laa,sug,rul,gri}. At last, letting high $T_c$ compounds go normal requires a huge, often unpractical magnetic field\cite{and,rul1,laa,sug,rul,gri}. 		
	\end{itemize}	
	Consequently, despite countless published $H_c$ data\cite{par,tin,sch,gen,and,rul1,laa,sug,rul,gri,ande,arm}, there is still no theory of the superconducting to normal transition, apart from the phenomenological\cite{par,tin,sch,gen} approach, based on the grossly wrong assumption $E_b=\mu_0H_c^2/2$. Thus the purpose of the present article is to design one, valid for \textit{both} high and low $T_c$ superconductors, as well. Accordingly, since it has been argued recently\cite{sz4} that feeding a growing current into a superconductor drives \textit{continuously} the superconducting phase to the normal one, this article will focus on a theoretical account of the \textit{current-driven}, superconducting to normal transition. Such an experimental procedure enables one to dodge all of the shortcomings mentioned above, in particular because reliable data for the critical current are available in all superconductors, including those for which $H_c$ is so large that it cannot be reached experimentally.  Furthermore it allows for a \textit{quantitative} treatment, unlike the $H$ mediated procedure. At last, since the current, carried by the superconducting electrons, plays a paramount role hereafter, it is worth mentioning an original view\cite{koi1}, which establishes the common significance of the persistent currents\cite{koi2} and Josephson effect\cite{jos}.\par
	The purpose of this work is then twofold :
\begin{itemize}
	\item 
	this transition will be studied with help of Newton's law and thermodynamics;
		\item 
	 the resulting findings will be taken advantage of to shed light into the  transport properties of high $T_c$ compounds in the controversial\cite{ande,arm,zan,led} $T>T_c$ range and the magneto-elastic behaviour, observed in elementary superconductors\cite{ols,ols2} for $T\leq T_c$.
\end{itemize}\par
	 The outline is as follows : the electrodynamical and thermodynamical properties of the superconducting to normal transition are worked out in sections $2,3$, respectively; a new derivation of the BCS calculation is given in section $4$, which enables us to define two critical temperatures and to reckon the $T$ dependent critical current; this analysis is further applied to investigate the transport properties of high-$T_c$ compounds for $T>T_c$, in section $5$; magneto-elastic data\cite{ols,ols2} are discussed  in section $6$; the results of this work are summarized in the conclusion.
		\section{Electrodynamical discussion}
	 As done previously\cite{sz4,sz2,sz1,sz3}, our analysis will proceed within the two-fluid model. Accordingly, the conduction electrons make up a homogeneous mixture of normal and superconducting electrons, in concentration $c_n,c_s$, respectively. The normal electrons behave like a Fermi gas\cite{ash}, characterised by $T$ and the Fermi energy $E_F$. The Helmholz free energy of independent electrons per unit volume $F_n$ and $E_F$ are related\cite{ash,lan} by $E_F=\frac{\partial F_n}{\partial c_n}$. By contrast, the superconducting electrons are organised as a many bound electron state\cite{sz4} of eigenenergy per unit volume $\mathcal{E}_s(c_s)$, such that its chemical potential reads $\mu=\frac{\partial \mathcal{E}_s}{\partial c_s}$.	Gibbs and Duhem's law\cite{lan} entails that the thermal equilibrium is characterised by
\begin{equation}
\label{gidu}
E_F(T,c_n(T))=\mu(c_s(T))\quad,
\end{equation}
with $c_0=c_n(T)+c_s(T)$ and $c_0$ being the concentration of conduction electrons.\par
\begin{figure}
\includegraphics*[height=6 cm,width=6 cm]{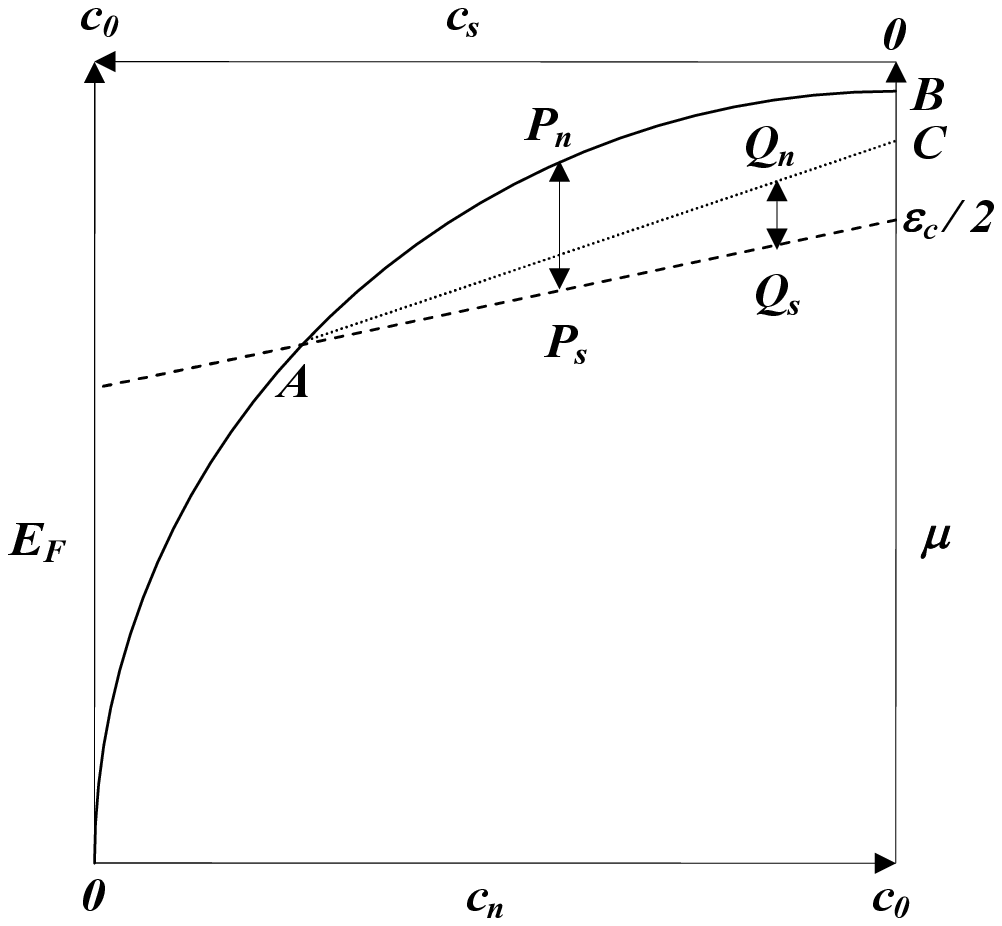}
\caption{schematic plots of $E_F(T<T_c,c_n)$ and $\mu(c_s)$ as solid and dashed lines, respectively; the origin $E_F=\mu=0$ is set at the bottom of the conduction band; the tiny difference $E_F(T,c_n)-\mu(c_0-c_n)$ has been hugely magnified for the reader's convenience; the crossing point $A$ between the solid and dashed lines represents the electron system in thermal equilibrium at $T,T_*$, according to Eq.(\ref{gidu}); the isothermal  process, addressed in sections $2,3$ is pictured by the $P_n,P_s$ pair, whereas the dotted line and the $Q_n,Q_s$ pair illustrate the adiabatic process, discussed in section $5$}\label{iso}
\end{figure}
	Consider then a superconducting material of cylindrical shape, characterized by its symmetry axis $z$ and radius $r_0$ in a cylindrical frame with coordinates ($r,\theta,z$) and flown through, along the $z$ direction, by a time dependent current $I(t)=\pi r_0^2j(t)$, with $j(t)$ being a uniform current density. The analysis of an isothermal, current-driven, superconducting to normal transition, outlined elsewhere\cite{sz4}, will be developed below with $j(t)=\gamma t, \gamma>0$. Accordingly, the initial state of the whole electron system is defined as $j(0)=0,c_n=c_n(T),c_s=c_s(T)$ (see $A$ in Fig.\ref{iso}). As $j(t)$ increases at constant $T$, the electron system shifts away from the equilibrium position in $A$ : the Fermi gas, represented by $P_n$ in Fig.\ref{iso}, moves, along the solid line, towards $B$, corresponding to the normal state $c_n=c_0$, while the superconducting electrons, represented by $P_s$, go, along the dashed line, towards the point characterized\cite{sz4} by $\mu(c_s=0)=\frac{\varepsilon_c}{2}$ ($\varepsilon_c$ refers to the Cooper pair energy\cite{cooper}). As this process will be shown to be reversible, the pair $P_n,P_s$, will shift back along the solid and dashed lines and will eventually merge into $A$, if $j$ is brought back down to $0$. The rest of section $1$   below deals with a detailed, quantitative account of the isothermal process, outlined above and illustrated in Fig.\ref{iso}.\par
		Due to $\gamma=\frac{dj}{dt}\neq 0$, Newton's law reads\cite{sz4,sz2,sz1,sz3} for the normal and superconducting current densities $j_n(t)$, $j_s(t)$ ($\Rightarrow j=j_n+j_s$; {note that $j_s$ is also referred\cite{koi1} to as a \textit{collective mode} current})
\begin{equation}
\label{newt}
\tau_n\frac{dj_n}{dt}=\sigma_n E-j_n\quad,\quad\tau_s\frac{dj_s}{dt}=\sigma_s\left(E-E_{s\rightarrow n}\right)-j_s.
\end{equation}
$E$ and $\tau_n,\tau_s$ are, respectively, the applied electric field and the decay times of $j_n,j_s$, due to friction with the lattice, responsible for Ohm's law, whereas $\sigma_n=\frac{c_ne^2\tau_n}{m},\sigma_s=\frac{c_se^2\tau_s}{m}$ stand for the normal and superconducting conductivities\cite{sz1,sz3} ($\tau_n<<\tau_s\Rightarrow\sigma_n<<\sigma_s$) and $m,e$ refer to the effective\cite{ash} mass and charge of an electron. Moreover $\tau_s$ being finite has been demonstrated elsewhere\cite{sz3} and shown\cite{sz4} furthermore to be consistent with observation of persistent currents at $E=0$. The effective field $E_{s\rightarrow n}$ is defined with respect to $f_{s\rightarrow n}=c_seE_{s\rightarrow n}$, the interelectron force, which turns superconducting electrons into normal ones. The resulting $f_{s\rightarrow n}$ is mediated by the interelectron coupling, also responsible for the binding energy of the superconducting electrons, i.e. $E_b>0$. Actually $E_{s\rightarrow n}$ was neglected in previous\cite{sz4,sz2,sz1,sz3} works ($\Rightarrow j_s\approx\sigma_sE$). But, as it will appear below that $\left|\frac{E_{s\rightarrow n}}{E}\right|<<1$, such an approximation was fully vindicated.\par
	During the elementary time-duration $\delta t$, superconducting electrons in concentration $\delta c_s$, moving at the mass center velocity\cite{sz4,sz2,sz1,sz3} $v_s$ ($\Rightarrow v_s=\frac{j_s}{c_se}$), are driven normal at vanishing velocity by $f_{s\rightarrow n}$, which corresponds to a momentum variation per unit volume of $\delta p=-m\delta c_sv_s$. Thence $f_{s\rightarrow n}$ is inferred from Newton's law to read
\begin{equation}
\label{newt2}
f_{s\rightarrow n}=\frac{\delta p}{\delta t}=-\frac{m\dot{c}_s}{c_se}j_s\Rightarrow
E_{s\rightarrow n}=-\frac{m\dot{c}_s}{(c_se)^2}j_s
\quad,
\end{equation}
with $\dot{c}_s=\frac{dc_s}{dt}$. Then combining Eqs.(\ref{newt},\ref{newt2}), while recalling that the inertial force $\propto \frac{dj_s}{dt}$ is negligible\cite{sz2,sz1}, yields
\begin{equation}
\label{ohm}
\begin{array}{c}
E=\frac{j_n}{\sigma_n}=\frac{j_s}{\sigma_s}+E_{s\rightarrow n}\Rightarrow
\frac{j_n}{\sigma_n}=\frac{j_s}{\sigma_s^*}=\frac{j}{\sigma_n+\sigma_s^*}\\
 \sigma_s^*=\frac{\sigma_s}{1-\tau_s\frac{\dot{c}_s}{c_s}}
\end{array}\quad .
\end{equation}
Eq.(\ref{ohm}) conveys the same meaning as Ohm's law, written for $j_n,j_s$ flowing parallel to each other, except for the effective conductivity $\sigma_s^*$ showing up  instead of $\sigma_s$.\par
	The elementary work $\delta W$, needed for one superconducting electron, moving with velocity $v_s$, to go normal with vanishing velocity, is reckoned to be equal to $\delta W=\frac{mv_s^2}{2}=\frac{m}{2}\left(\frac{j _s}{c_se}\right)^2$, thanks to the kinetic energy theorem. On the other hand, for an isothermal process, $\delta W$ is also equal to the difference of free energy\cite{lan} between the superconducting and normal states, which leads thence to the identity\cite{lan} $\delta W=\frac{\partial F_n}{\partial c_n}-\frac{\partial \mathcal{E}_s}{\partial c_s}=E_F(T,c_n)-\mu(c_s)$. Consequently, $j_s$ reads finally
\begin{equation}
\label{js}
j_s(c_s)=c_se\sqrt{\frac{2}{m}\left(E_F(T,c_0-c_s)-\mu(c_s)\right)}\quad.
\end{equation}
Note that, unlike the normal current $j_n=\sigma_nE$, $j_s$ is \textit{independent} from the external field $E$ and depends only on the \textit{concentration} of bound electrons $c_s$.\par
	Eq.(\ref{ohm}) can now be recast as an ordinary differential equation of first order for the unknown $c_s(j)$
\begin{equation}
\label{ed2}
\gamma\frac{d\log c_s}{dj}=\frac{\frac{c_s}{c_n\tau_n}\left(\frac{j}{j_s(c_s)}-1\right)-\frac{1}{\tau_s}}{\frac{c_s}{c_n}\left(\frac{j}{j_s(c_s)}-1\right)-1}\quad ,
\end{equation}
with $c_n=c_0-c_s$ and $j_s$ given by Eq.(\ref{js}).\par
\begin{figure}
\includegraphics*[height=6 cm,width=6 cm]{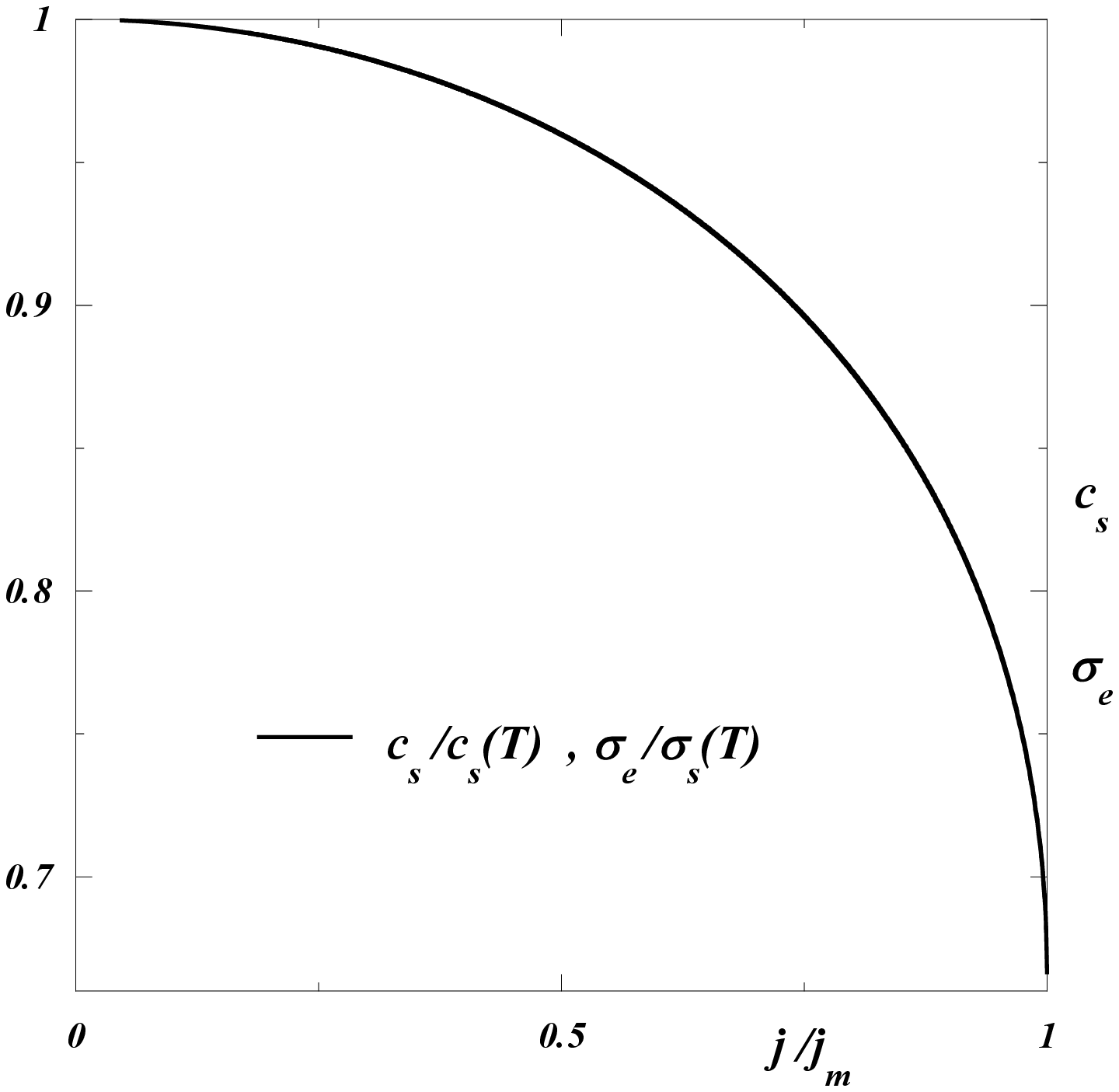}
\caption{plots of $c_s(j),\sigma_e(j)$ for $j\leq j_m$, as a solid line; the results have been found to be independent from $\gamma$ in accordance with Eq.(\ref{js1})}
\label{jm3}
\end{figure}
\begin{figure}
\includegraphics*[height=13 cm,width=6 cm]{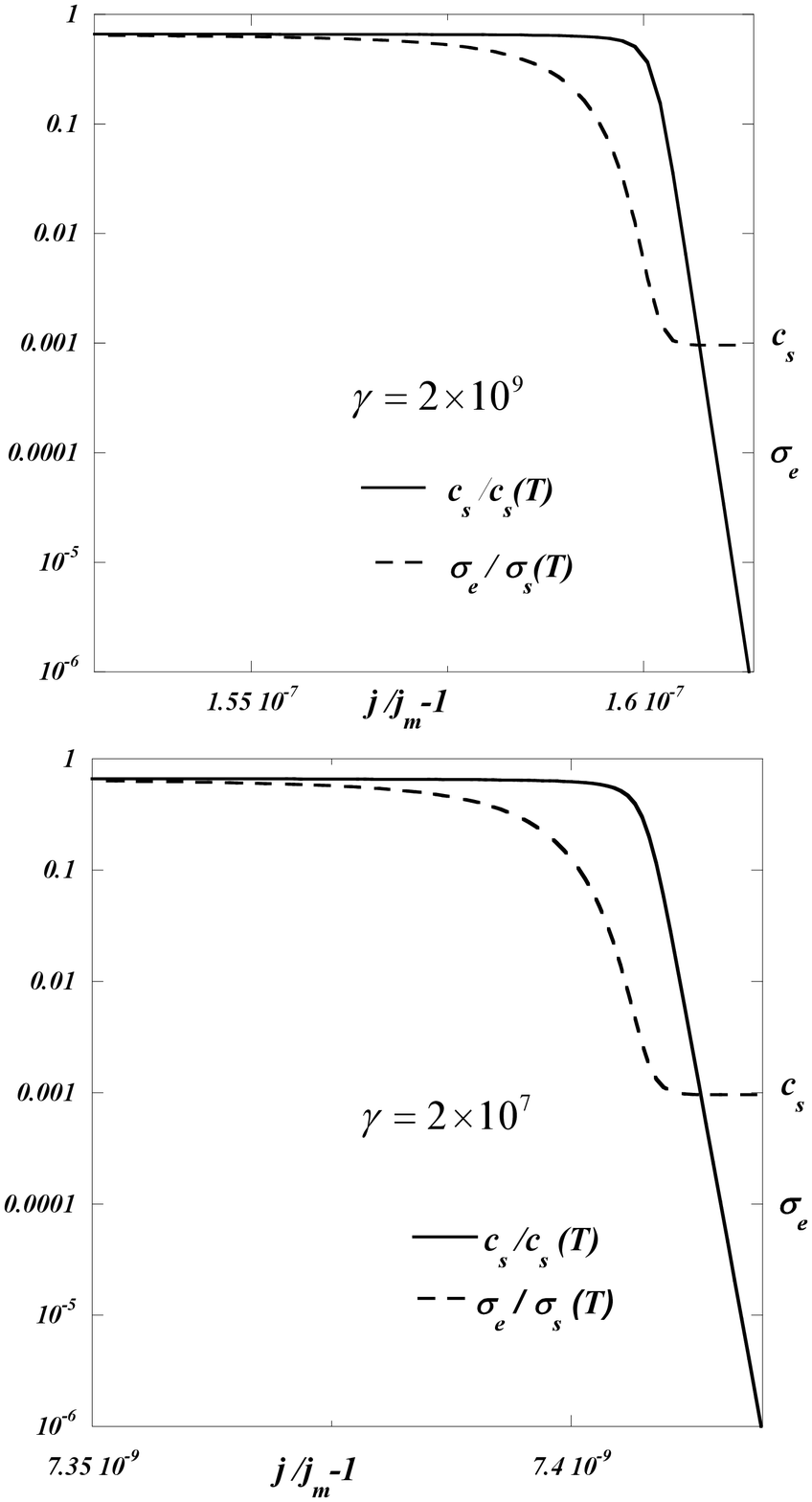}
\caption{plots of $c_s(j),\sigma_e(j)$ for $j> j_m$, as a solid and dashed line, respectively; the calculation has been done for two values $\gamma=2\times 10^{7},2\times 10^{9}A/(m^2\times s)$}
\label{jm4}
\end{figure}

	For $j$ increasing from $0$, $c_s$ decreases from $c_s(T)$, while $E_F-\mu$ increases from $0$, proportionally to the length of the arrow linking $P_n,P_s$ in Fig.\ref{iso}. In addition, since $j_s$ will eventually vanish for $c_s\rightarrow0$, as inferred from Eq.(\ref{js}), $j_s$ is bound to rise from $j_s=0$ at $A$ up to a maximum $j_s(c_m)$ at $c_m(T)<c_s(T)$ defined by $\frac{dj_s}{dc_s}(c_m)=0$. In order to solve Eq.(\ref{ed2}), $E_F-\mu$ will be replaced by its Taylor's expansion at first order with respect to $c_s-c_s(T)$
\begin{equation}
\label{js2}
\frac{e^2}{m}\left(E_F-\mu\right)=\beta\left(c_s(T)-c_s\right)\Rightarrow 
c_m=\frac{2}{3}c_s(T)\quad,
\end{equation}
with $\beta=\frac{e^2}{m}\left(\frac{\partial E_F}{\partial c_n}(c_n(T))+\frac{\partial \mu}{\partial c_s}(c_s(T))\right)$. Thus Eq.(\ref{ed2}) has been integrated with $c_0=10^{28}/m^3,c_s(T)=.1c_0,\tau_s=10^{-9}s,\tau_n=10^{-4}\tau_s,\beta=10^{-65}A^2\times m^5$ and initial condition $c_s(j=0)=c_s(T)$. The resulting data $c_s(j),\sigma_e(j)$  ($\sigma_e=\sigma_n+\sigma_s^*$ refers to the effective conductivity) have been plotted in Figs.\ref{jm3},\ref{jm4}, corresponding to $j\leq j_m$ or $j>j_m$, respectively, with $j_m$ defined by $j_m=j_s(c_m)$.\par 
	For $j\leq j_m$, there is $\tau_s\left|\frac{\dot{c}_s}{c_s}\right|<<1$, so that Eq.(\ref{ohm}) yields $\sigma_s^*\approx\sigma_s$ and Eq.(\ref{ed2}) reduces to 
\begin{equation}
\label{js1}
\frac{j_s(c_s)}{j}=1+\frac{\sigma_n}{\sigma_s}=1+\frac{\tau_n}{\tau_s}\left(\frac{c_0}{c_s}-1\right)\quad. 
\end{equation}
 Likewise, Eq.(\ref{js1}) implies $j_s\approx\sigma_sE\Rightarrow \left|\frac{E_{s\rightarrow n}}{E}\right|<<1$, which confirms the validity of a previous assumption. As $\gamma$ does not show up in Eq.(\ref{js1}), there is a one-to-one correspondence between $j$ and $c_s$, as seen in Fig.\ref{jm3}. Moreover, $\tau_n<<\tau_s$ entails that $j_s\approx j$, so that the $c_s(j),\sigma_e(j)$ plots cannot be distinguished from each other. Note that $\frac{dc_s}{dj}(j=0)=0$, while $\left|\frac{dc_s}{dj}\right|$ becomes very large for $j\rightarrow j_m$.\par
	However, when $j$ keeps growing beyond $j_m$, $j_s\approx j$ is no longer valid because of $j_s\leq j_m<j$. Consequently, as seen in Fig.\ref{jm4} , $c_s(j)$, obtained by integrating Eq.(\ref{ed2}) for $j>j_m$, falls steeply from $c_s(j_m)=c_m$ down to $0$, and  $\sigma_e$ sinks by the ratio $\frac{c_s(T)\tau_s}{c_0\tau_n}=10^3$ from $\sigma_e(T)\approx\sigma_s(T)$ down to $\sigma_e(T_c)=\frac{c_0e^2\tau_n}{m}$, typical of the normal metal. Meanwhile $j$ undergoes a tiny increase from $j_m$ up to $j_M$, with $j_M$ being weakly $\gamma$ dependent, i.e. $j_M/j_m -1\approx 10^{-7},10^{-8}$ for $\gamma=2\times 10^9,2\times 10^7 A/(m^2\times s)$, respectively (see Fig.\ref{jm4}). Finally, due to $j_M\approx j_m = j_s(c_m)$, $j_M(T<T_c)$ reads 
$$j_M=ec_m(T)\sqrt{\frac{2}{m}\left(E_F\left(T,c_0-c_m(T)\right)-\mu(c_m(T))\right)}\quad .$$
\par
	Integrating Eq.(\ref{ed2}) from $j=j_M$ down to $j=0$  with the initial condition $c_s=c_s(j_M)\approx 0$, while keeping $\gamma$ \textit{unaltered}, will produce the same solution $c_s(j)$, as displayed in Figs.\ref{jm3},\ref{jm4}. This shows that the superconducting-normal transition is \textit{reversible} and there is a \textit{one-to-one} correspondence between $j$ and $c_s$, provided that $\gamma$ keeps the  \textit{same} value for $j$ increasing from $0$ up to $j_M$ or decreasing from $j_M$ down to $0$, as well. This property holds actually for any $j(t)$, such that $j(t)=j(t_p-t), \forall t\in\left[0,t_p/2\right]$, with $t_p$ taken such that $j(t_p/2)=j_M$.\par
	Due to $j_s\approx j$ for $j<j_m$, measuring $\sigma_e(j)$ and the $j$ dependent London length, which gives access\cite{sz1,sz3} to $c_s$, would enable one to chart $E_F(T,c_n)-\mu(c_s)$ with help of Eq.(\ref{js}). Given the highest observed $j_M$ values, Eq.(\ref{js}) provides the estimate $E_F(T,c_n)-\mu(c_s)<10^{-5}eV$. It is noticeable that the conductivity, decreasing by several orders of magnitude for $j\rightarrow j_M$, as seen above, and for $T\rightarrow T_c^-$, as discussed elsewhere\cite{sz3}, is to be ascribed, in both cases, to $c_s$ decreasing very steeply down to $0$.\par
		\section{Thermodynamical discussion}
	As recalled above, the work $W_{s\rightarrow n}$, performed by $f_{s\rightarrow n}$, whereby all of superconducting electrons are turned into normal ones via an isothermal process\cite{lan}, is equal to the difference of free energy per unit volume $\Delta F$, between the normal and superconducting states. Due to the very definitions of $E_F,\mu$, the work $W_{s\rightarrow n}$ is thence deduced to read  
\begin{equation}
\label{w1}
W_{s\rightarrow n}=\int_0^{c_s(T)}\left(E_F\left(T,c_0-u\right)-\mu\left(u\right)\right)du\quad .  
\end{equation}
In addition, Eq.(\ref{w1}) implies that $W_{s\rightarrow n}=-W_{n\rightarrow s}$, consistently with the reversible nature of the transition.\par
	$W_{s\rightarrow n}$ can be also achieved alternatively by using the definition of $\Delta F=\Delta\mathcal{E}+T\Delta S$, with $\mathcal{E},S$ being, respectively, the total energy and entropy of the sample, i.e. including all of the \textit{lattice} and \textit{electron} degrees of freedom. $\Delta\mathcal{E},\Delta S$ will be calculated by working out the detailed thermal balance over the following trajectory : the sample is first taken at $T<T_c$ and heated up to $T_c$ with $j=0$. Hence, the associated $\Delta\mathcal{E}_1,\Delta S_1$ read
\begin{equation}
\label{df1}
\begin{array}{l}
\Delta\mathcal{E}_1=\int_{T}^{T_c}\left(C_\phi(u)+C_s(u)\right)du\\
\Delta S_1=\int_{T}^{T_c}\left(C_\phi(u)+C_s(u)\right)\frac{du}{u}
\end{array}
\quad ,  
\end{equation}
with $C_\phi(T),C_s(T)$ standing for the respective contributions\cite{ash} to the specific heat of the phonons (Debye) and of the conduction electrons in the superconducting state; then let the sample be cooled down back to $T$, while being flown through by a current density $j\geq j_M(T)$, so that the sample remains normal down to $T$. The associated $\Delta\mathcal{E}_2,\Delta S_2$ read then   
\begin{equation}
\label{df2}
\begin{array}{l}
\Delta\mathcal{E}_2=\int_{T_c}^{T}\left(C_\phi(u)+C_n(u)\right)du\\
\Delta S_2=\int_{T_c}^{T}\left(C_\phi(u)+C_n(u)\right)\frac{du}{u}
\end{array}
\quad ,  
\end{equation}
with $C_n(T)$ standing for the $T$ linear, specific heat of a Fermi gas\cite{ash}, which is known to be independent from $j$, like $C_\phi(T)$. At last, the searched expressions read\cite{gor}
\begin{equation}
\label{df3}
\begin{array}{l}
E_b(T)=\Delta\mathcal{E}_1+\Delta\mathcal{E}_2=\int_{T}^{T_c}\left(C_s(u)-C_n(u)\right)du\\
W_{s\rightarrow n}(T)=\Delta\mathcal{E}_1+T\Delta\mathcal{S}_1+\Delta\mathcal{E}_2+T\Delta\mathcal{S}_2\\
\quad \quad \quad \quad \quad =\int_{T}^{T_c}\left(C_s(u)-C_n(u)\right)\left(1-\frac{T}{u}\right)du
\end{array}
,  
\end{equation}
with $E_b(T)$ being the binding energy of the superconducting phase with respect to the normal one at $T$. Noteworthy is that the superconducting phase being stable ($\Leftrightarrow E_b(T)>0$) requires $C_s(T)>C_n(T)$ in Eq.(\ref{df3}), which is confirmed experimentally\cite{par,ash}, i.e. $C_s(T_c)\approx 3C_n(T_c)$.\par
	$W_{s\rightarrow n}$ can actually be measured directly by feeding a growing current $I(t)=\pi r_0^2\gamma t$ into the superconducting sample, from $t=0$ until $t=\frac{t_p}{2}$ with $I(\frac{t_p}{2})=\pi r_0^2 j_M(T)$, so that the sample goes normal at $\frac{t_p}{2}$ (this is referred to as the Silsbee effect\cite{par}). Then $I(t)$ is reduced, like $I(t)=\pi r_0^2\gamma\left(t_p-t\right)$, from $I\left(\frac{t_p}{2}\right)$ down to $I(t_p)=0$. The work $W(t_p)$, performed by the electric field $E$ from $t=0$ until $t=t_p$, reads then
\begin{equation}
\label{w2}
\begin{array}{c}
W(t_p)=W_1+W_2\\
W_1=\int_0^{\frac{t_p}{2}}U(t)I(t)dt\quad ,\quad W_2=\int^{t_p}_{\frac{t_p}{2}}U(t)I(t)dt
\end{array}\quad ,
\end{equation}
with $U=El$ and $l$ being the measured voltage drop across the sample and its length, respectively. Moreover, owing to Eq.(\ref{ohm}), $W_1,W_2$ can be recast as
\begin{equation}
\label{w4}
\begin{array}{c}
\frac{W_1}{\pi r_0^2l}=\int_0^{\frac{t_p}{2}}\left(\frac{j_n^2}{\sigma_n}+\frac{j_s^2}{\sigma_s}+j_sE_{s\rightarrow n}\right)dt\\
\frac{W_2}{\pi r_0^2l}=\int_{\frac{t_p}{2}}^{t_p}\left(\frac{j_n^2}{\sigma_n}+\frac{j_s^2}{\sigma_s}+j_sE_{n\rightarrow s}\right)dt
\end{array}\quad .
\end{equation}
Likewise, recalling that $j_n,\sigma_n,j_s,\sigma_s$ have been shown above to depend on $j$ only, if $\gamma$ is kept fixed, and furthermore
$$\begin{array}{c}
W_{s\rightarrow n}=\int_0^{\frac{t_p}{2}}j_sE_{s\rightarrow n}dt\quad,\quad W_{n\rightarrow s}=\int_{\frac{t_p}{2}}^{t_p}j_sE_{n\rightarrow s}dt\\
W_{s\rightarrow n}=-W_{n\rightarrow s}
\end{array}\quad ,$$
enables us to recast $W_1,W_2$ as
\begin{equation}
\label{w6}
\begin{array}{c}
\frac{W_1}{\pi r_0^2l}=Q_1+W_{s\rightarrow n}\quad ,\quad \frac{W_2}{\pi r_0^2l}=Q_1-W_{s\rightarrow n}\\
Q_1=\int_0^{j_M(T)}\left(\frac{j_n^2}{\sigma_n}+\frac{j_s^2}{\sigma_s}\right)\frac{dj}{\gamma}
\end{array}\quad ,  
\end{equation}
with $Q_1$ expressing the Joule heat, released\cite{sz4} through process I per unit volume  for $t\in\left[0,t_p/2\right]$. Finally it ensues from Eq.(\ref{w6}) 
\begin{equation}
\label{w5}
\frac{W_1-W_2}{2\pi r_0^2l}=\int_{T}^{T_c}\left(C_s(u)-C_n(u)\right)\left(1-\frac{T}{u}\right)du\quad .  
\end{equation}
The validity of Eq.(\ref{w5}) should be checked experimentally first in a superconducting material, for which accurate data are available for $C_s,C_n$ and thence $T_c$ is low\cite{ash} enough for $C_s>C_\phi, C_n>C_\phi$. Accordingly, $Al$ ($T_c=1.19 K$) might be a good candidate. In case of a successful test, Eq.(\ref{w5}) might then provide with a rather \textit{unique} access to $C_s$ in high $T_c$ materials, for which the direct measurement of $C_s$ proves unreliable\cite{lor}, due to $C_s<<C_\phi$. Note that $C_n$ can always be measured at low $T$ by feeding into the sample a current density $j>j_M(T)$, whereby the sample goes normal even at $T<T_c$, because $C_n$ is $j$ independent, unlike $C_s$, and extrapolated further to higher $T$, by taking advantage of its $T$ linear behaviour\cite{ash}.\par
	Although the superconducting to normal transition and ice melting into water are both first order processes, they differ in two respects :
\begin{itemize}
	\item 
	the role of the latent \textit{heat}, typical of all usual first order transitions (melting or vaporisation), is played here by the latent \textit{work} $W_{s\rightarrow n}$, because the superconducting-normal transition is controlled by \textit{current}  rather than by \textit{temperature}; 
		\item 
	ice and water are separated by a clear-cut interface, whereas the mixture of superconducting and normal electrons is homogeneous. Consequently, the chemical potentials of ice and water remain uniquely defined all over the melting process, while $E_F,\mu$ vary continuously with $c_s\in\left[0,c_s(T<T_c)\right]$.
\end{itemize}
	\section{Critical current}
As shown elsewhere\cite{sz4}, the \textit{bound electron} current $j_s$ will turn out to be \textit{persistent}, only if the necessary condition $\sigma_s+\sigma_J>0$, with $\sigma_J<0$ being the conductivity characterising process II of the Joule effect, is fulfilled. It conveys the physical meaning that the negative Joule heat, released via the anomalous process II, typical of superconductors, should prevail over the positive one, stemming from the regular process I. Since $\sigma_J$ reads\cite{sz4} as $\sigma_J=-\frac{(ev)^2\tau_s}{\left|\frac{\partial \mu}{\partial c_s}\right|}$ with $mv^2\approx 1eV$, the inequality $\sigma_s+\sigma_J>0$ can be recast as
	\begin{equation}
\label{pre}
 r(c_s )=\frac{c_s}{mv^2}\left|\frac{\partial \mu}{\partial c_s}\right|>1\quad .  
\end{equation}
As it will appear below that $\frac{\partial \mu}{\partial c_s}$ remains finite for $c_s\rightarrow 0$, the inequality (\ref{pre}) is bound not to hold any more for $c_s<c_c$ with the critical concentration $c_c$ defined by $r(c_c)=1$. To proceed further, $c_c$ must be assessed, which requires to reckon $\frac{\partial \mu}{\partial c_s}$. The only practical tool for this purpose is the BCS scheme\cite{bcs}, but for some reason to become clear below, we shall refrain from using it, and rather develope our own procedure.\par
	Thus let us consider a three-dimensional crystal containing $N$ sites and $2n$ itinerant electrons with $N>>1,n>>1$ ($\Rightarrow c_s=2n/N$). These electrons of spin $\sigma=\pm1/2$ populate a single band, accomodating at most two electrons per site ($\Rightarrow n\leq N$). The independent electron motion is described, in reciprocal space, by the Hamiltonian $H_d$
\begin{equation} 
\label{hd}
H_d=\sum_{k,\sigma}\varepsilon(k)c^+_{k,\sigma}c_{k,\sigma}\quad ,
\end{equation} 
for which $\varepsilon(k),k$ are the one-electron, spin-independent energy ($\Rightarrow\varepsilon(k)=\varepsilon(-k)$) and a vector of the Brillouin zone, respectively, and the sum over $k$ is to be carried out over the whole Brillouin zone. Then $c^+_{k,\sigma},c_{k,\sigma}$ are  one-electron creation and  annihilation operators on the Bloch state $\left|k,\sigma\right\rangle$ 
	$$\left|k,\sigma\right\rangle=c^+_{k,\sigma}\left|0\right\rangle\quad ,\quad \left|0\right\rangle=c_{k,\sigma}\left|k,\sigma\right\rangle\quad ,$$
with $\left|0\right\rangle$ being the no electron state. They enable us to introduce the two-electron creation and annihilation operators\cite{bcs,cooper,ja1} $b^+_{K,k}=c^+_{k,+}c^+_{K-k,-},b_{K,k}=c_{K-k,-}c_{k,+}$, {which operate on hard-core bosons and thence do not fulfill the boson commutation rules}. The interacting electron motion is  governed by a truncated Hubbard Hamiltonian $H_K$, used previously\cite{bcs,ja1} 
\begin{equation} 
\label{hK}
H_K=\sum_{k}\varepsilon(K,k)b^+_{K,k}b_{K,k}+\frac{U}{N}\sum_{k,k'}b^+_{K,k}b_{K,k'}\quad,  
\end{equation} 
with $\varepsilon(K,k)=\varepsilon(k)+\varepsilon(K-k)$ and $U$ being the Hubbard coupling constant. Unlike previous authors\cite{tin,sch,bcs,cooper}, we shall consider \textit{both} cases $U>0,U<0$.\par
	The eigenstate of the Schr\"{o}dinger equation, pertaining to a  single bound pair, $\left(H_K-\varepsilon_c(K)\right)\varphi_c=0$, is known as the Cooper pair\cite{cooper} state $\varphi_c=\sum_{k}\frac{b^+_{K,k}}{\varepsilon_c(K)-\varepsilon(K,k)} \left|0\right\rangle$, with the eigenenergy $\varepsilon_c(K)$ being the solution of 
\begin{equation} 
\label{coo}
\frac{1}{U}=\frac{1}{N}\sum_{k}\frac{1}{\varepsilon_c(K)-\varepsilon(K,k)}=\int_{-t_K}^{t_K}\frac{\rho_K(\varepsilon)}{\varepsilon_c(K)-\varepsilon}d\varepsilon.  
\end{equation} 
$\pm t_K$ are the upper and lower bounds of the two-electron band, i.e. the maximum and minimum of $\varepsilon(K,k)$ over $k$, whereas $\rho_K(\varepsilon)$ is the corresponding two-electron density of states. For the sake of illustration, we shall solve Eq.(\ref{coo}) for $\rho_K(\varepsilon)=\frac{2}{\pi t_K}\sqrt{1-\left(\frac{\varepsilon}{t_K}\right)^2},t_K=t\cos\left(Ka/2\right)$, where $t,a$ are the one-electron bandwidth and the lattice parameter, respectively. The dispersion curves $\varepsilon_c(K)$ are given in Fig.4 for $U>0$ only, because it can be deduced from Eq.(\ref{coo}) and $\rho_K(\varepsilon)=\rho_K(-\varepsilon)$ that $\varepsilon_c(K,U)=-\varepsilon_c(K,-U)$. A remarkable feature  is that $\varepsilon_c(K)\rightarrow t_K$, i.e. the upper bound of the two-electron band, for $U$ decreasing toward $t_K/2$, so that there is \textit{no} Cooper pair solution for $U<t_K/2$ (accordingly, the dashed curve is no longer defined in Fig.4 for $\frac{Ka}{\pi}<.13$), in marked contrast with the opposite conclusion reached elsewhere\cite{cooper}, that there is a Cooper pair, even for $U\rightarrow 0$. This discrepancy results from the three-dimensional Van Hove singularities, showing up at both two-electron band edges $\rho_K\left(\varepsilon\rightarrow\pm t_K\right)\propto \sqrt{t_K-\left|\varepsilon\right|}$, unlike the two-electron density of states, used previously\cite{cooper}, which displayed no such singularity. 	\par
\begin{figure}
\label{coop}
\includegraphics*[height=6 cm,width=6 cm]{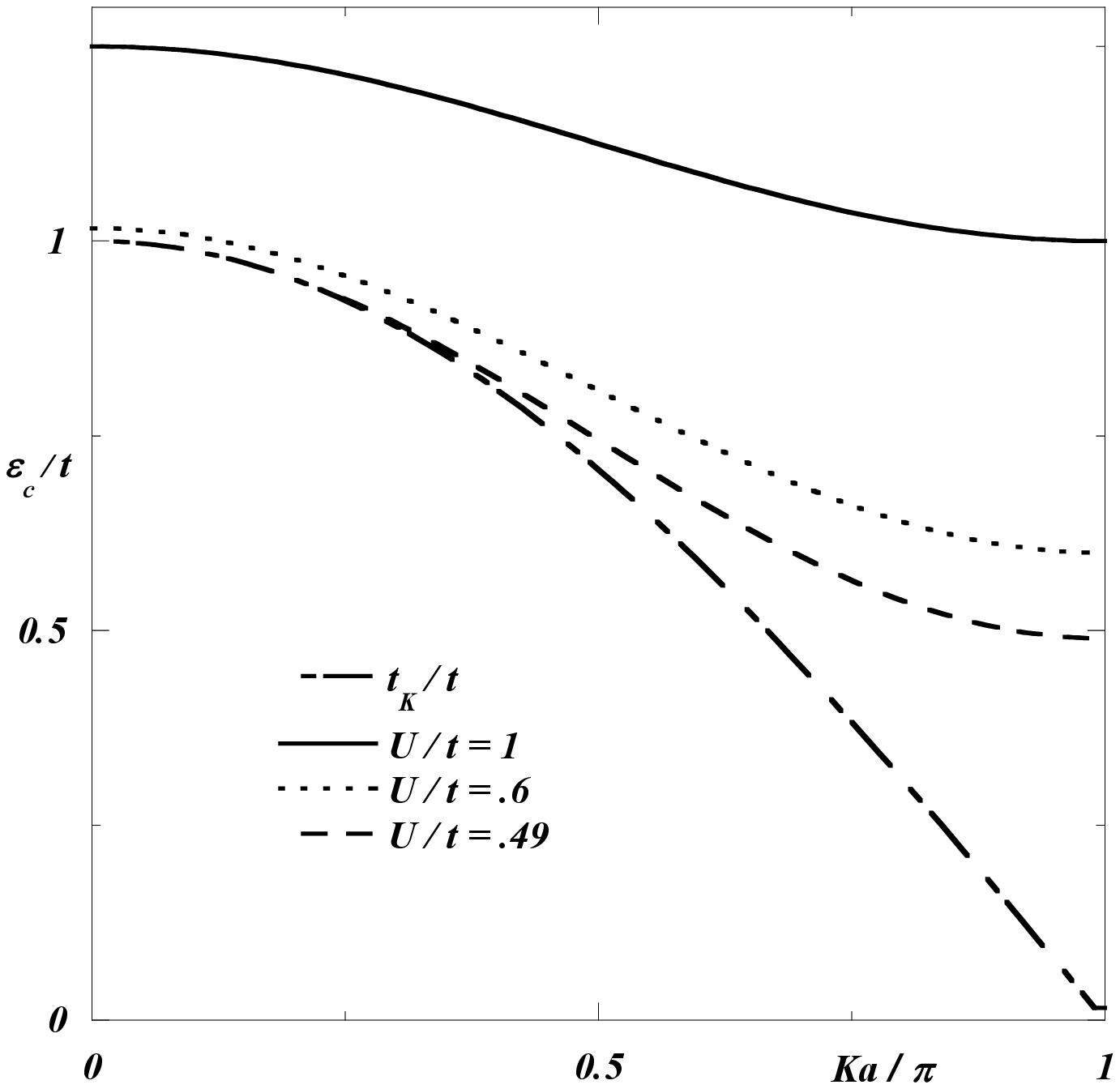}
\caption{dispersion curves of $t_K$ as a dashed-dotted line and of $\varepsilon_c(K)$ as solid, dashed and dotted lines, associated with various $U$ values, respectively}
\end{figure}
	$H_K$ operates within the Hilbert space $S_K$. A typical vector of its basis reads $\varphi=\prod_{i=1,..m}b^+_{K,k_i}\left|0\right\rangle$ with $m$ being any integer. We shall look for a variational approximation $\varphi_v$ of the single\cite{ja2,ja3} bound eigenstate of $H_K$ inside the subset $\left\{\varphi_v\right\}\subset S_K$, characterized by 
$$\left\langle \varphi_v\right|b^{(+)}_{K,k}\left|\varphi_v\right\rangle=\left(n_k\left(1-n_k\right)\right)^\alpha,\quad\forall\varphi_v\in\left\{\varphi_v\right\}\quad.$$ 
The real parameter $\alpha$ will be assigned shortly and $n_k=\left\langle \varphi_v\right|b^+_{K,k}b_{K,k}\left|\varphi_v\right\rangle$. The pair number operator $b^+_{K,k}b_{K,k}$ has two eigenvalues $0,1$, associated with $\left|0\right\rangle$ and $
b^+_{K,k}\left|0\right\rangle$, respectively, so that $0\leq n_k\leq 1$ and $\sum_kn_k=n$. The energy of $\varphi_v$ per site reads 
$$\begin{array}{c}
\mathcal{E}=\frac{\left\langle \varphi_v\right|H_K\left|\varphi_v\right\rangle}{N}=\sum_{k}\varepsilon(K,k)\frac{n_k}{N}+U\Delta^2\\
\Delta=\sum_{k}\frac{\left(n_k\left(1-n_k\right)\right)^\alpha}{N}
\end{array}\quad.$$ 
Hence minimising $\mathcal{E}$ ($\Rightarrow d\mathcal{E}=0$), under the constraint of $n$ kept constant ($\Rightarrow dn=0$), yields
$$\varepsilon(K,k)+2\alpha U\Delta\frac{1-2n_k}{\left(n_k\left(1-n_k\right)\right)^{1-\alpha}}=\lambda\quad,$$ 
with $\lambda=\frac{\partial\mathcal{E}}{\partial n_k},\forall k$ being a Lagrange multiplier, which implies that $\lambda=\frac{\partial\mathcal{E}}{\partial n}=2\mu(c_s=\frac{2n}{N})$. The $\alpha$ value will be assigned now by checking consistency with the Cooper pair properties in the limit $n\rightarrow 0\Rightarrow n_k\rightarrow 0\Rightarrow n_k\propto\left(2\mu-\varepsilon(K,k)\right)^{\frac{1}{\alpha-1}}$. Comparing with $n_k=\left\langle \varphi_c\right|b^+_{K,k}b_{K,k}\left|\varphi_c\right\rangle$ $\propto\left(\varepsilon_c(K)-\varepsilon(K,k)\right)^{-2}$, inferred from Eq.(\ref{coo}), yields finally $\alpha=1/2$ and $\varepsilon_c(K)=2\mu(c_s=0)$, a conclusion which had already been reached by an independent rationale\cite{sz4}. Hence, our variational procedure can be summarised, with help of notations introduced elsewhere\cite{tin}, as follows 
\begin{equation} 
\label{bcs2}
\begin{array}{c}
\tan 2\theta_k=\frac{2U\Delta}{2\mu-\varepsilon(K,k)}\quad,\quad  n_k=\sin ^2\theta_k\\
\Delta=\sum_{k}\frac{\sin 2\theta_k}{2N}=\int_{-t_K}^{t_K}\frac{\sin 2\theta(\varepsilon)}{2}\rho_K(\varepsilon)d\varepsilon\\
 c_s=2\sum_{k}\frac{\sin ^2\theta_k}{N}=2\int_{-t_K}^{t_K}\sin ^2\theta(\varepsilon)\rho_K(\varepsilon)d\varepsilon
\end{array}\quad,  
\end{equation} 
with $0\leq\theta_k\leq\frac{\pi}{2} $. The formulae in Eqs.(\ref{bcs2}) are found to be identical to those of BCS\cite{tin,sch,bcs}. As an illustrative example, Eqs.(\ref{bcs2}) have been solved for $\mu(c_s),\Delta(c_s)$ and $r(c_s )$ defined by Eq.(\ref{pre}), with $U>0$ and $c_s<1$ electron per site, only, because it can be deduced from Eqs.(\ref{bcs2}) and $\rho_K(\varepsilon)=\rho_K(-\varepsilon)$ that $\mu(c_s,U)=-\mu(c_s,-U)$$=-\mu(2-c_s,U)$. The results, presented in Fig.5, exhibit $\frac{\partial \mu}{\partial c_s}$ almost independent from $c_s$ and $\Delta(c_s\rightarrow0)\propto\sqrt{c_s}$. An important inequality, holding for $U>0$ and $U<0$ as well, can be deduced from $\mu(c_s,U)=-\mu(c_s,-U)$ and Fig.5
\begin{equation} 
\label{udmu}
			U\frac{\partial\mu}{\partial c_s}<0
\quad.  
\end{equation}\par 
Note that for $K=\frac{\pi}{a}$, the two-electron band is dispersionless because of $t_{K=\frac{\pi}{a}}=0$. Then applying Eqs.(\ref{bcs2}) to the $K=\frac{\pi}{a}$ case gives $\Delta=\frac{\sin 2\theta}{2}$ and finally $\mu=\frac{U}{2}\left(1-c_s\right)$, the validity of which can be checked independently, because the $n$-pair, bound eigenstate of $H_{K=\frac{\pi}{a}}$ is known\cite{ja1} to read $\varphi=\sum_{i=1,..d}\frac{\varphi_i}{\sqrt{d}}$, with $d=\left(\begin{array}{c}N\\n\end{array}\right)$, $\varphi_i=\prod_{l=1,..n}b^+_{\frac{\pi}{a},k_{i_l}}\left|0\right\rangle$ and the sum with respect to $i$ runs over all of $n$ pair combinations $\left\{b^+_{\frac{\pi}{a},k_{i_1}},...b^+_{\frac{\pi}{a},k_{i_n}}\right\}$, chosen among $N$ of available pairs. As each $\varphi_i$ contributes $\frac{n}{dN}(1-\frac{n}{N})U$ to $\mathcal{E}$, it can thence be inferred $\mathcal{E}=U\frac{c_s}{2}(1-\frac{c_s}{2})\Rightarrow\mu=\frac{\partial\mathcal{E}}{\partial c_s}=\frac{U}{2}\left(1-c_s\right)$, which is seen to be identical to the above result, deduced from Eqs.(\ref{bcs2}). At last, there is $U\frac{\partial\mu}{\partial c_s}=-U^2/2<0$ in 	accordance with inequality (\ref{udmu}).\par
\begin{figure}
\label{5}
\includegraphics*[height=13 cm,width=6 cm]{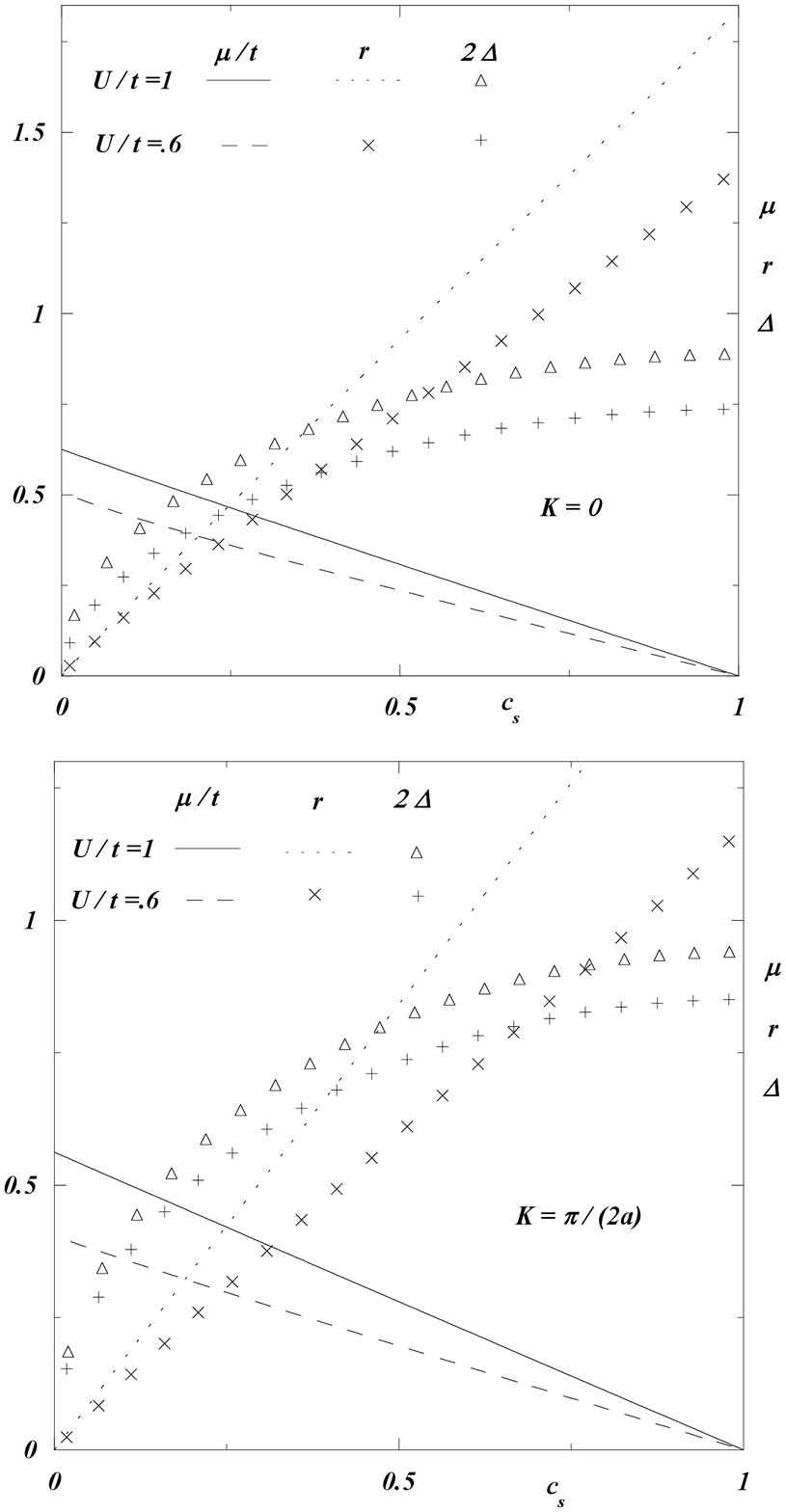}
\caption{plots of $\mu(c_s),\Delta(c_s),r(c_s)$ reckoned for $K=0,\frac{\pi}{2a}$; the solid and dotted lines and the triangles, which pertain to $\mu,r,\Delta$, respectively, have been calculated with $U/t=1$, whereas the dashed line and the $\times,+$ symbols, which refer to $\mu,r,\Delta$, respectively, have been calculated with $U/t=.6$; the $r$ data have been calculated with $mv^2=t/3$; $c_s=1$ corresponds to one electron per site}
\end{figure}
	Combining Eq.(\ref{coo}) with Taylor's expansions of $\sin2\theta_k$, $\sin^2\theta_k$, worked out from Eqs.(\ref{bcs2}), up to $\Delta^2$ for $c_s\rightarrow 0\Rightarrow\Delta\rightarrow 0$, leads to
	$$\begin{array}{l}
	\frac{1}{U}=\int_{-t_K}^{t_K}\frac{\rho_K(\varepsilon)}{\varepsilon_c(K)-\varepsilon}d\varepsilon\\
	=\int_{-t_K}^{t_K}\left(1-2\left(\frac{U\Delta}{2\mu-\varepsilon}\right)^2\right)\frac{\rho_K(\varepsilon)}{2\mu-\varepsilon}d\varepsilon\\
	\left(U\Delta\right)^2=\frac{c_s}{2\int_{-t_K}^{t_K}\frac{\rho_K(\varepsilon)}{\left(\varepsilon_c(K)-\varepsilon\right)^2}d\varepsilon}
	\end{array}\quad .$$
Subtracting the integrals equal to $1/U$ from each other, while taking advantage of $\mu(c_s\rightarrow 0)\rightarrow\varepsilon_c(K)/2$, gives in turn
	$$\left(U\Delta\right)^2=\left(\frac{\varepsilon_c(K)}{2}-\mu\right)\frac{\int_{-t_K}^{t_K}\frac{\rho_K(\varepsilon)}{\left(\varepsilon_c(K)-\varepsilon\right)^2}d\varepsilon}{\int_{-t_K}^{t_K}\frac{\rho_K(\varepsilon)}{\left(\varepsilon_c(K)-\varepsilon\right)^3}d\varepsilon}
	\quad .$$
Equating both expressions of $\left(U\Delta\right)^2$ yields finally
	$$\frac{\partial\mu}{\partial c_s}(K,c_s=0)=-\frac{\int_{-t_K}^{t_K}\frac{\rho_K(\varepsilon)}{\left(\varepsilon_c(K)-\varepsilon\right)^3}d\varepsilon}{2\left(\int_{-t_K}^{t_K}\frac{\rho_K(\varepsilon)}{\left(\varepsilon_c(K)-\varepsilon\right)^2}d\varepsilon\right)^2}
		\quad .$$
Note that $\left|\frac{\partial\mu}{\partial c_s}(K,c_s=0)\right|\rightarrow\infty$ for $U\rightarrow t_K/2$ and\\
 $U\frac{\partial\mu}{\partial c_s}(K,c_s=0)<0$ in accordance with inequality (\ref{udmu}).\par
	With help of Fig.5, $c_c$, defined by $r(c_c )=1$ in Eq.(\ref{pre}), can now be assigned, for $K=0$, to the values $.7,.54$, and for $K=\frac{\pi}{2a}$, to the values $.85,.6$, associated with $U/t=.6,1$, respectively. Noteworthy is that $\varphi_v$ will sustain persistent currents or not, according to whether $c_s>c_c$ or $c_s<c_c$, although $\varphi_v$ undergoes \textit{no qualitative} change for $c_s=c_c$. Accordingly, it still obeys Eqs.(\ref{bcs2}) for $c_s>c_c$ and $c_s<c_c$, as well. Therefore, $\varphi_v(c_s<c_c)$ will be referred to below, as the many-bound-electron, \textit{non-superconducting} (MBENS) state. Moreover, applying Eq.(\ref{js}) for $c_s=c_c$, while taking advantage of the Sommerfeld integral\cite{ash} and Eq.(\ref{gidu}), yields the critical current density as
	\begin{equation}
	\label{jc}
	\begin{array}{l}
	j_c(T)=c_ce\sqrt{\frac{2}{m}\left(E_F\left(T,c_0-c_c\right)-\mu(c_c)\right)}\\
	\quad\quad\quad=\pi k_Bec_c\sqrt{\frac{\rho^{'}(E_F^*)}{3m\rho(E_F^*)}\left(T_*^2-T^2\right)}	
	\end{array}
	\quad,
	\end{equation}
with $E_F^*=E_F\left(T_*,c_0-c_c\right)$. $k_B,\rho(\varepsilon)$ designate Boltzmann's constant and the one-electron density of states and $\rho^{'}(\varepsilon)=\frac{d\rho}{d\varepsilon}$, while $T_*<T_c$ is defined by $E_F(T_*,c_0-c_c)=\mu(c_c)\Rightarrow c_s(T_*)=c_c$. The calculated behavior $j_c(T)\propto\sqrt{T_*^2-T^2}$, resulting from Eq.(\ref{jc}), is found to agree with observation\cite{tin,sch,gen,ash}. Consequently, the MBENS state and the superconducting one can be observed for $T_*<T<T_c$ and $T<T_*$, respectively. In a low $T_c$ metal such as $Sn$, $c_s$ has been shown\cite{sz3} to grow steeply from $c_s(T_c)=0$ up to $c_s(T_c-.04K)\approx c_s(0)\Rightarrow T_c-T_*<.04K$, so that $T_c,T_*$ are unlikely to be resolved experimentally from each other. However $T_c-T_*$ will be argued in the next section to be quite sizeable in high $T_c$ materials.\par 
		$\frac{\partial\mu}{\partial c_s}<0$ has been shown\cite{sz4} to be a prerequisite	for persistent currents. Hence, the inequality \ref{udmu} entails $U>0$. Besides, an additional setback of the assumption\cite{bcs,cooper} $U<0\Rightarrow\frac{\partial\mu}{\partial c_s}>0$ is to preclude any thermal equilibrium for $T<T_c$. Here is a proof by contradiction. Let us assume that the BCS state is indeed in equilibrium at $T_c$, which implies $E_F(T_c,c_0)=\mu(c_s=0)$, because of $c_s(T_c)=0, c_n(T_c)=c_0$, in accordance with Eq.(\ref{gidu}). When $T$ decreases from $T_c$ down to $0$, charge conservation $c_0=c_n(T)+c_s(T)$ entails
	$$E_F(T)=E_F(T_c)-\frac{c_s(T)}{\rho(E_F(T_c))},\quad
		\mu(T)=\mu(T_c)+\frac{\partial\mu}{\partial c_s}c_s(T),$$
for which we have used\cite{ash} $\frac{\partial E_F}{\partial c_s}=1/\rho(E_F)$, while neglecting $\frac{\partial E_F}{k_B\partial T}\approx k_BT/E_F<<1$. Thus, $\frac{\partial\mu}{\partial c_s}>0$ implies $E_F(T)<E_F(T_c)=\mu(T_c)<\mu(T)$, so that the equilibrium condition $\mu(c_s(T))=E_F(T,c_0-c_s(T))$ in Eq.(\ref{gidu}) cannot be fulfilled for any $T<T_c$. \textit{Q.E.D.}
	\section{High-$T_c$ compounds}
	Overdoped high $T_c$ compounds are known\cite{and,rul1,laa,sug,rul,gri,ande,arm,zan,led} to undergo, at $T_c$, a crossover from a superconducting state of type II, observed for $T<T_c$, to an ill-understood state, which sustains no persistent current, but the conduction properties of which differ yet markedly from those of usual metals up to $T>>T_c$ :
\begin{itemize}
	\item 
	contrary to the conductivity expected to be low, given the high doping rate $>.15$, it is observed to be large;
	\item
	the Hall coefficient is found to be $T$ dependent, which hints at a $T$ dependent carrier concentration, unlike what is observed in usual metals and alloys, behaving like a Fermi gas\cite{ash} with $T$ independent concentration.
\end{itemize}
Assuming $T_c=T_*$, both above mentioned features might be consistent with an electron system, comprising a Fermi gas \textit{and} a MBENS state in respective concentration $c_n(T>T_*),c_s(T)$ and fulfilling Eq.(\ref{gidu}) with $c_0=c_n(T)+c_s(T)$. As a matter of fact, $\tau_s>>\tau_n$ entails\cite{sz1} that the large conductivity is settled by the MBENS electrons only and the Hall coefficient, dominated by $j_s$, is $T$ dependent as is $c_s(T)$. The main virtue of such an assumption is that it lends itself to an experimental check, as shown below.\par
		Consider a thermally isolated sample, flown  through by $I(t)=\pi r_0^2j(t)$ with $j(t)=\gamma t$, and taken at $t=0$ in the thermal equilibrium state, represented by $A$ in Fig.\ref{iso}, i.e. $T(t=0)=T_*, I(t=0)=0,c_n(t=0)=c_n(T_*),c_s(t=0)=c_0-c_n(T_*)$. While $I(t)$ keeps growing, the bound electrons, pictured by $Q_s$ in Fig.\ref{iso}, are turned into independent ones, depicted by $Q_n$, as explained in section $2$. The experiment ends up at $t=t_f$, when $Q_n$, after traveling all along the dotted line, merges with $C$, referring to the normal state and thence characterized by $T(t_f)=T_f,c_n(t_f)=c_0,c_s(t_f)=0$. Thus applying the first law of thermodynamics to this adiabatic process yields
	\begin{equation}
	\label{adi}
	\begin{array}{c}
	\int_{T_*}^{T_f}C_\phi(T)dT=Q_1+Q_2\quad,\quad Q_2=\int_{0}^{t_f}\frac{j_s^2(t)}{\sigma_J}dt\\
	Q_1=\int_{0}^{t_f}\frac{U(t)}{l}j(t)dt=\int_{0}^{t_f}\left(\frac{j_n^2(t)}{\sigma_n}+\frac{j_s^2(t)}{\sigma_s}\right)dt
	\end{array},
	\end{equation}
where $Q_1>0,Q_2<0$ stand for the Joule heat released through processes\cite{sz4} I and II, respectively, and $C_n(T>T_*)<<C_\phi(T)$, $C_s(T>T_*)<<C_\phi(T)$, $W_{s\rightarrow n}<<Q_1$ have been neglected. Derivating Eq.(\ref{adi}) with respect to $t$ gives finally 
	\begin{equation}
	\label{adi1}
		C_\phi(T)\dot T=\frac{U(t)}{l}j(t)+\frac{j_s^2}{\sigma_J}=\frac{j_n^2}{\sigma_n}+j_s^2\left(\frac{1}{\sigma_s}+\frac{1}{\sigma_J}\right),
	\end{equation}
with $\dot T=\frac{dT}{dt}$. Because of $\frac{1}{\sigma_s}+\frac{1}{\sigma_J}>0$, due to the very definition of $T_*$, we predict $\dot T>0\Rightarrow T_f>T_*$, with $\sigma_J<0\Rightarrow C_\phi(T)\dot T<\frac{U(t)}{l}j(t)$ . Despite $\dot T>0$ like in a usual metal, the latter inequality would rather read $C_\phi(T)\dot T=\frac{U(t)}{l}j(t)$, if the same experiment were carried out in a normal conductor. Conversely, would the experiment be performed at $T<T_*\Rightarrow\frac{1}{\sigma_s}+\frac{1}{\sigma_J}<0$, we should observe\cite{sz4} $\dot T<0$, as remarked by De Gennes too (see\cite{gen} footnote in p.18). Besides, the sign of $\dot T$ is independent of that of $\dot I$, because the Joule effect is \textit{irreversible}. At last, due to the high doping rate, the local electron concentration is likely to display spatial fluctuations, which should eventually result into a sample, comprising both superconducting \textit{and} MBENS domains. This case could be brought to experimental evidence by observing different values of $\dot T$ in Eq.(\ref{adi1}), according to whether a \textit{dc} ($\Rightarrow\dot T_{dc}$) or \textit{ac} ($\Rightarrow\dot T_{ac}$) current is fed into the sample, because superconducting domains will contribute to the Joule effect \textit{only} for \textit{ac} current, whereas MBENS ones will do in \textit{both} cases. Thus we predict $\dot T_{ac}<\dot T_{dc}$.
	\section{Magnetoelasticity}
	Magnetoelastic effects were reported\cite{ols,ols2} long ago, in superconducting metals, at $T\leq T_c$ and atmospheric pressure : when the magnetic field $H$ starts growing from $0$, the sample first expands by a tiny amount ($\approx 10^{-7}$) and then shrinks abruptly for $H$ reaching some critical value $H_c(T)$, at which the sample goes normal. Actually, because the superconducting electrons are known\cite{par,tin,sch,gen} to be in a macroscopic singlet spin state, $H$ has no direct sway on them, but merely induces an eddy current according to Faraday's law\cite{sz2}. This current, responsible for the Meissner effect, turns superconducting electrons into normal ones, as discussed in section $2$, but only within a thin film of thickness $\lambda_M$, located at the outer edge of the sample\cite{sz2}. Meanwhile, the partial pressure, stemming from the electrons, is altered, as will be shown now.\par
	The free energy, associated with a sample of volume $V$, containing $n$ conduction electrons ($\Rightarrow \frac{n}{V}=c_0=c_n+c_s$), reads $VF(T,c_0)$ with $F(T,c_0)=F_n(T,c_n)+\mathcal{E}_s(c_s)$ being the electronic free energy per unit volume. The partial pressure $p_e$, exerted by the electrons, reads\cite{lan} then
	\begin{equation}
	\label{p1}
	\begin{array}{l}
	p_e(H\neq 0)=-\frac{\partial\left(VF\right)}{\partial V}=c_0\frac{\partial F}{\partial c_0}-F\\
	\quad\quad=c_nE_F(T,c_n)-F_n+c_s\mu(c_s)-\mathcal{E}_s	
	\end{array}\quad,
	\end{equation}
with $c_n>c_n(T),F_n=\int_0^{c_n}E_F(T,u)du$, $c_s=c_0-c_n,\mathcal{E}_s=\int_0^{c_s}\mu(u)du$. \par
	Eq.(\ref{p1}) implies
 $$\frac{\partial p_e}{\partial c_n}=c_n\frac{\partial E_F}{\partial c_n}-c_s\frac{\partial \mu}{\partial c_s}\quad.$$
Besides, $\frac{\partial E_F}{\partial c_n}=\frac{1}{\rho(E_F)},\frac{\partial \mu}{\partial c_s}<0$ entail $\frac{\partial p_e}{\partial c_n}>0$. Since $c_n$ grows at the expense of $c_s$ for increasing $H$, the inequality $\frac{\partial H}{\partial c_n}>0$ is always valid, which implies at last $\frac{\partial p_e}{\partial H}>0$, in agreement with the observed $H$ induced expansion\cite{ols,ols2}.\par
	For $H=H_c(T)$, the sample goes normal, so that $H$ penetrates suddenly into bulk matter and polarises the whole set of normal electrons in concentration $c_0$. The associated paramagnetic energy per unit volume reads\cite{ash} $\mathcal{E}_H=-\frac{\left(\mu_BH\right)^2}{2}\rho(E_F(T,c_0))$ with $\mu_B$ being the Bohr magneton. Because Pauli's susceptibility is $T$ independent\cite{ash}, $\mathcal{E}_H$ is also equal to the magnetic contribution to the free energy, so that the partial pressure $p_H$, associated with $H$, reads
 $$p_H=c_0\frac{\partial \mathcal{E}_H}{\partial c_0}-\mathcal{E}_H=\frac{\left(\mu_BH\right)^2}{2}\left(\rho(E_F)-c_0\frac{\rho'(E_F)}{\rho(E_F)}\right)\quad,$$
with $E_F=E_F(T,c_0)$. As the sample was reported\cite{ols,ols2} to shrink at $H_c(T)$, this implies $p_H<0$, which can be realized only if $E_F(T,c_0)$ lies close to a Van Hove singularity at $\varepsilon_{VH}\Rightarrow\rho'(E_F)\propto \left(E_F-\varepsilon_{VH}\right)^{-1/2}>>1$.\par
	This kind of $H$ driven experiment provides merely qualitative information, because of several drawbacks, related to the Meissner effect\cite{sz2}, as recalled in section $1$. Consequently, the critical field $H_c(T)$ is ill-defined. To buttress this conclusion, we shall calculate $H(r)$ induced by the homogeneous current density $j_c$, parallel to the $z$ axis. $H(r)$ is normal\cite{sz2} to the unit vectors along the $r$ and $z$ coordinates and there is $H=rj_c/2$, thanks to the Amp\`ere-Maxwell equation. Hence, $H$ is seen to vary from $H(r=0)=0$ up to $H(r_0)=r_0j_c/2$, so that $H_c$ cannot be defined in a \textit{unique} way, unlike $j_c(T)$. Likewise superconductors of type II make this proof more cogent, inasmuch as the whole superconducting sample is known\cite{par,tin,sch,gen} to turn continuously normal over a \textit{broad} range of critical values $H_c\in\left[H_{c_1},H_{c_2}\right]$ with $H_{c_1}<<H_{c_2}$.\par
	\section{Conclusion}
	 A unified picture, accounting for low and high $T_c$ superconductivity as well, has been developed. The physical significance of two different critical temperatures $T_*,T_c$ with $T_*<T_c$, characterizing the electrodynamical behavior of superconducting materials, has been analyzed. Whereas no persistent current can be observed for $T>T_*$, $T_c$ is the upper bound of the MBENS state ($\Leftrightarrow c_s(T\geq T_c)=0$) and is also identical to the usual critical temperature. The expression of the maximum persistent current $j_c(T)$ has been worked out and found to agree with observation. Unlike the normal current $j_n$, the bound electron current $j_s$ does not depend on the applied electric field, but rather on $c_s$. The many-body wave-function, describing the motion of bound electrons, is identical for both superconducting ($c_s>c_c$) \textit{and} MBENS ($c_s<c_c$)  states, and accurately approximated by the BCS variational scheme. Conversely, the critical field $H_c$ has been shown to lack a unique definition. Whereas $T_*,T_c$ are unlikely to be resolved from each other in conventional superconductors due to the steep variation of $c_s(T\rightarrow T_c)$, $T_c/T_*$ may be $>10$ in high $T_c$ compounds. Moreover, their peculiar conduction properties in the contentious\cite{ande,arm,zan,led} range $T\in\left[T_*,T_c\right]$ have been ascribed to a MBENS state and an experiment, taking full advantage of the interplay between the usual and anomalous Joule effects\cite{sz4}, has been outlined to check the validity of this assumption. The merit of a current driven experiment over a $H$ driven one has been emphasized. At last, it has been shown that a \textit{repulsive} ($U>0$) Hubbard coupling is a prerequisite for superconductivity at thermal equilibrium, in accordance with the Coulomb force and Eq.(\ref{gidu}).

\end{document}